\documentclass[%
aps,pre,reprint,%
showpacs,showkeys,%
raggedbottom%
]{revtex4-1}
\usepackage{amsfonts,amssymb,amsmath}
\usepackage{graphicx}
\usepackage{xcolor}
\usepackage{subfigure}
\usepackage{hyperref}

\begin{document}

\title{Fractional Langevin Equation of Distributed Order}

\author{C. H. Eab}
\email[]{chaihok.E@chula.ac.th}
\affiliation{Department of Chemistry, Faculty of Science,
             Chulalongkorn University,
             Bangkok 10330, Thailand
}
\author{S. C. Lim}
\email[]{sclim47@gamail.com}
\thanks{corresponding author}
\affiliation{52, Jalan Desa Maju,
Taman Desa,
58100 Kuala Lumpur,
Malaysia
}

\date{\today}

\begin{abstract}
Distributed order fractional Langevin-like equations are introduced and applied to describe anomalous diffusion without unique diffusion or scaling exponent.
It is shown that these fractional Langevin equations of distributed order can be used to model the kinetics of retarding subdiffusion whose scaling exponent decreases with time, 
and the strongly anomalous ultraslow diffusion with mean square displacement which varies asymptoically as a power of logarithm of time.
\end{abstract}


\maketitle
\section{Introduction}
\label{sec:introduction}
Anomalous transport phenomena occur in many physical systems. 
Particles in complex media, 
instead of the normal diffusion they undergo anomalous diffusion. 
The basic property commonly used to characterize different types of diffusion is the mean-squared displacement (MSD) of the diffusing particles. 
The diffusion is known as anomalous diffusion if the MSD is no longer linear in time as in normal diffusion or Brownian motion; 
instead it satisfies a power law behavior and varies as $t^\alpha$, $0 < \alpha < 2$. 
MSD with $\alpha <1$  corresponds to subdiffusion, 
which is slower than the normal diffusion. 
It is known as superdiffusion when $\alpha > 1$ with the diffusion faster than the normal one.
Just like normal diffusion which can be described by diffusion equation and Langevin equation, 
it is possible to model anomalous diffusion using fractional version of these equations. 
The stochastic processes commonly associated with anomalous diffusion include fractional Brownian motion and Levy motion 
\cite{MetzlerKlafter00,KalgesRadonsSokolov08,SamorodnitskyTaqqu94}.
In order to describe anomalous diffusion, 
the usual random walk model for normal diffusion has to be replaced by the continuous-time random walk model 
\cite{MetzlerKlafter00,GorenfloMMPP02,GorenfloVivoliMainardi04}.

Anomalous diffusion forms only a portion of diffusion processes in nature. 
There also exist many diffusion processes which do not have MSD that varies as power law with a unique diffusion or scaling exponent $\alpha$. 
For such systems $\alpha$  is no longer a constant, instead it can be a function of certain physical parameters like position, time, temperature, density, etc. 
Kinetic equations of constant fractional order such as fractional diffusion equation and fractional Langevin equation are successful in describing anomalous diffusion. 
However, for diffusion process with non-unique scaling exponent, 
constant order fractional kinetic equations are not applicable. 
It is necessary to introduce kinetic equations of multi-fractional order or variable fractional order 
\cite{KobelevKobelevKlimontovich03,ChechkinGorenfloSokolov05,UmarovSteinberg09}
We remark that variable order fractional differential equations are mathematically intractable except for few very simple cases. 
On the other hand, there is a subclass of diffusion processes which have different regimes with distinct scaling exponents. 
For example, single-file diffusion of Brownian particles confined to narrow channels or pores 
\cite{KalgesRadonsSokolov08,KargerRuthven92,LimTeo09}.
There also exists ultraslow diffusion such as the diffusion observed in Sinai model 
\cite{Sinai82} 
which has MSD that varies asymptotically as power of logarithm of time.
It is possible to model such diffusion processes with a different type of multifractional differential equation based on distributed order derivative first introduced by Caputo 
\cite{Caputo67}.

Since its introduction in the 1960s by Caputo \cite{Caputo67}, 
distributed order differential equation was subsequently developed by Caputo himself 
\cite{Caputo01,Caputo03}, 
and also by Bagley and Torvik 
\cite{BagleyTorvik00a,BagleyTorvik00b}.
Various authors 
\cite{LorenzoHartley02,%
ChechkinGorenfloSokolov02,%
ChechkinGSG02,%
ChechkinKlafterSokolov03,%
Naber04,%
SokolovChechkinKlafter04a,%
SokolovChechkinKlafter04b,%
Atanackovic05,%
GorenfloMainardi05,%
MeerschaertScheffler06,%
GorenfloMainardi06,%
Hanyga07,%
MainardiMGS07,%
Kochubei08,%
MainardiMPG08,%
AtanackovicPilipovicZorica09a,%
AtanackovicPilipovicZorica09b%
}
have applied fractional differential equations of distributed order to model 
anomalous, non-exponential relaxation processes and anomalous diffusion with non-unique diffusion or scaling exponent.
 
The distributed order time derivative is defined by
\begin{align}
D_{(\varphi)} f(t) & = \int_{\beta _1}^{\beta _2}
                       \varphi\left(\alpha\right) {D^\alpha} f\left(t\right)d\alpha
\label{eq:introduction_00010}
\end{align}
where the weight function $\varphi(\alpha)$  is a positive integrable function defined on $[\beta_1,\beta_2]$.
For our purpose, it is assumed that $0 \leq \beta_1 < \alpha < \beta_2 \leq 1$.
Here the fractional time derivative $D^\alpha$ can be either the Riemann-Liouville or Caputo type 
\cite{Podlubny99, MetzlerKlafter00,KilbasSrivastavaTrujillo06},
which are defined respectively for $n-1 < \alpha < n$ as
\begin{align}
 D_{RL}^\alpha f(t) & = \frac{1}{\Gamma (n - \alpha )}
                      \frac{d^n}{dt^n}
                      \int_0^t \frac{f(u)du}{(t - u)^{\alpha  - n + 1}} 
\label{eq:introduction_00020}
\end{align}
and
\begin{align}
 D_{C}^\alpha f(t) & = \frac{1}{\Gamma(n - \alpha )}
                      \int_0^t \frac{f^{(n)}(u)du}{(t - u)^{\alpha  - n + 1}}.
\label{eq:introduction_00030}
\end{align}
For our purpose, we let $\beta_1=0$ and $\beta_2=1$ in the subsequent discussion. 

Due to the fact that distributed order derivative modifies the fractional order of derivative by integrating all possible orders over a certain range, 
the solution of the resulting fractional equation is not characterized by a definite scaling exponent. 
One can regard the distributed order fractional time derivatives as time derivatives on various time scales. 
Derivatives with distributed fractional order can be used to describe transport phenomena in complex heterogeneous media with multifractal property. 
Such processes exhibit memory effects over various time scales. 
Distributed order time-fractional diffusion equation and space-fractional diffusion equation have been considered by various authors 
\cite{%
ChechkinGorenfloSokolov05,%
UmarovSteinberg09,%
KargerRuthven92,%
LimTeo09,%
Caputo03,%
BagleyTorvik00a,%
BagleyTorvik00b,%
LorenzoHartley02,%
ChechkinGorenfloSokolov02%
}.
Solutions of these equations can be used to describe retarding subdiffusion and accelerating subdiffusion, 
as well as accelerating superdiffusion 
\cite{%
ChechkinGorenfloSokolov02,%
ChechkinGSG02,%
SokolovChechkinKlafter04a,%
SokolovChechkinKlafter04b,%
MeerschaertScheffler06,%
MainardiMPG08%
}
and ultraslow diffusion 
\cite{%
ChechkinGorenfloSokolov02,%
ChechkinGSG02,%
ChechkinKlafterSokolov03,%
Naber04,%
GorenfloMainardi05,%
MeerschaertScheffler06,%
Hanyga07,%
Kochubei08,%
MainardiMPG08%
}.

Distributed order fractional diffusion equations of various types have been quite well-studied. 
On the other hand, distributed order Langevin equations have not been considered as far as we are aware. 
The main aim of this paper is to study the Langevin-like equations of distributed order and to consider their possible applications. 
This paper considers several types of fractional Langevin equation of distributed order. 
The statistical properties, in particular the MSD, of the solutions to these equations are studied. 
Possible applications of these equations to model retarding subdiffusion such as single-file diffusion 
and ultraslow diffusion such as in the Sinai model and some other systems will be discussed.

\section{Fractional Langevin of distributed order}
\label{sec:DOfracLamgevin}
Let us consider a simple free fractional Langevin equation without frictional term
\begin{align}
  {D^\alpha}x(t) & = \xi(t), \quad  n-1 < \alpha < n,
\label{eq:DOfracLamgevin_00010}
\end{align}
where $\xi(t)$ is a stationary Gaussian random noise with mean zero and covariance $C(\tau) = \left<x(t+\tau)x(t)\right>$
to be specified later on. 
Using the Laplace transform of Riemann-Liouville derivative $D_{RL}^\alpha f(t)$  and Caputo derivative $D_{C}^\alpha f(t)$ 
\cite{ChechkinGSG02,Naber04}:
\begin{align}
  \mathcal{L}\left[{D_{RL}^\alpha}f(t)\right](s) & = {s^\alpha}\tilde{f}(s) - \sum_{k=0}^{n-1} s^k\left[D_{RL}^{\alpha-k-1}f(t)\right]_{t=0}
\label{eq:DOfracLamgevin_00020}
\end{align}
and
\begin{align}
  \mathcal{L}\left[{D_{C}^\alpha}f(t)\right](s) & = {s^\alpha}\tilde{f}(s) - \sum_{k=0}^{n-1} s^{\alpha-k-1}f^{(k)}(0),
\label{eq:DOfracLamgevin_00030}
\end{align}
one gets the Laplace transform of 
(\ref{eq:DOfracLamgevin_00010}) 
 for the Riemann-Liouville and Caputo cases are respectively
 \begin{align}
   {s^\alpha}\tilde{x}(s) - \sum_{k=0}^{n-1} s^k\left[D_{RL}^{\alpha-k-1}x(t)\right]_{t=0}
           & = \tilde{\xi}(s)
\label{eq:DOfracLamgevin_00031}
 \end{align}
and
\begin{align}
  {s^\alpha}\tilde{x}(s) - \sum_{k=0}^{n-1} s^{\alpha-k-1}x^{(k)}(0) & = \tilde{\xi}(s).
\label{eq:DOfracLamgevin_00032}
\end{align}
The solution is
\begin{align}
  x(t) & = a(t) + \frac{1}{\Gamma(\alpha)}\int_0^t (t-u)^{\alpha-1}\xi(u) du 
\label{eq:DOfracLamgevin_00040}
\end{align}
where 
$a(t)$ is the inverse Laplace transform of   
$\displaystyle\sum_{k=0}^{n-1} s^k\left[D_{RL}^{\alpha-k-1}x(t)\right]_{t=0}$
and   
$\displaystyle\sum_{k=0}^{n-1} s^{\alpha-k-1}x^{(k)}(0)$
for the Riemann-Liouville and Caputo case respectively. 
If the random noise is white noise $\eta(t)$ with covariance  $\left<\eta(t)\eta(s)\right>=\delta(t-s)$, 
then the variance of the process is given by
\begin{align}
  \sigma^2(t) & = \frac{1}{\Gamma(\alpha)^2}\int_0^t (t - u)^{2\alpha-2} du
               = \frac{t^{2\alpha-1}}{(2\alpha-1)\Gamma(\alpha)^2},
\label{eq:DOfracLamgevin_00050}
\end{align}
and it is the same for both Caputo and Riemann-Liouville derivative. 
If one assumes $a(t)=0$, 
(\ref{eq:DOfracLamgevin_00040}) 
can be regarded as the definition of Riemann-Liouville fractional Brownian motion with the Hurst index
$H=\alpha - \frac{1}{2}$ 
\cite{SithiLim95,Lim01}.
When $\alpha=1$, we get Brownian motion with $H=1/2$.	

One can now generalize 
(\ref{eq:DOfracLamgevin_00010}) 
to the case of distributed order fractional Langevin equation
\begin{align}
  {D_{(\varphi)}}x(t) & = \xi(t), \quad t \geq 0,
\label{eq:DOfracLamgevin_00060}
\end{align}
with $D_{(\varphi)}$  as defined as in 
(\ref{eq:introduction_00010}).
Note that though 
(\ref{eq:DOfracLamgevin_00060}) 
is in the form of free fractional Langevin equation of distributed order, 
we shall see in next section that the distributed order derivative term will introduce ``frictional'' terms depending on the type of 
weight function $\varphi(\alpha)$. 
Laplace transform of 
(\ref{eq:DOfracLamgevin_00060}) 
is 
\begin{align}
  A(s)\tilde{x}(s) - B(s) & = \tilde{\xi}(s)
\label{eq:DOfracLamgevin_00070}
\end{align}
with
\begin{align}
  A(s) & = \int_0^1 \varphi(\alpha) {s^\alpha} d\alpha, \quad 0 \leq \alpha \leq 1,
\label{eq:DOfracLamgevin_00080}
\end{align}
\begin{subequations}
    \label{eq:DOfracLamgevin_00090}
  \begin{align}
    B(s) & = \int_0^1 d\alpha\varphi(\alpha) \left(\sum_{k=0}^{\lceil\alpha\rceil}
             s^k\left[D_{RL}^{\alpha-k-1}x(t)\right]_{t=0}\right)
\label{eq:DOfracLamgevin_00090a}
\end{align}
for Riemann-Liouville case, and 
\begin{align}
    B(s) & =  \int_0^1 d\alpha\varphi(\alpha) \left(\sum_{k=0}^{\lceil\alpha\rceil}
               s^{\alpha-k-1}x^{(k)}(0)\right) 
    \label{eq:DOfracLamgevin_00090b}
  \end{align}
\end{subequations}
for Caputo case.
Here $\lceil\alpha\rceil$ denotes the largest integer smaller or equal to $\alpha$.

Solving 
(\ref{eq:DOfracLamgevin_00070}) 
gives
\begin{align}
  \tilde{x}(s) & = \frac{B(s)}{A(s)} + \frac{\tilde{\xi}(s)}{A(s)},
\label{eq:DOfracLamgevin_00100}
\end{align}
with inverse Laplace transform
\begin{align}
  x(t) & = m(t) + \int_0^t G(t - \tau)\xi(\tau)d\tau,
\label{eq:DOfracLamgevin_00110}
\end{align}
where $m(t)$ and $G(t)$ are the inverse Laplace transform of $B(s)/A(s)$  and $1/A(s)$  respectively.	

The mean and variance of the process $x(t)$ are given by
\begin{align}
  \bar{x} & = \left<x(t)\right> = m(t),
\label{eq:DOfracLamgevin_00120}
\end{align}
and
\begin{align}
  \sigma^2(t) & = \left<\left(x(t) - \bar{x}\right)^2\right> 
                = \int_0^t\int_0^t G(u)C(u-v)G(v) dudv \nonumber \\
              & = 2\int_0^tG(u)\int_0^u C(u-v)G(v)dvdu
\label{eq:DOfracLamgevin_00130}
\end{align}
Thus, the solution to the distributed order fractional Langevin equation 
(\ref{eq:DOfracLamgevin_00060}) 
have the same variance for both the Riemann-Liouville and Caputo cases. 
Their means are different except when $x^{(k)}(0)$ and $\left[D_{RL}^{\alpha-k-1}x(t)\right]=0$, $k=0,1,\cdots,n$,
and in this case the MSD for the two type of fractional derivatives are the same and equal to the variance.
This is in contrast to time-fractional diffusion equation of distributed order, 
which leads to stochastic process with different MSD or variance for the Riemann-Liouville and Caputo case 
\cite{%
ChechkinGorenfloSokolov02,%
ChechkinGSG02,%
SokolovChechkinKlafter04a,%
SokolovChechkinKlafter04b,%
MeerschaertScheffler06,%
MainardiMPG08%
}.
In our subsequent discussion $D^\alpha$ and $D_{(\varphi)}$    shall denote respectively fractional derivative and distributed order fractional derivative of  either Riemann-Liouville or  Caputo type. The two terms MSD and variance shall be used interchangeably

Before we discuss various examples of distributed order Langevin equation, 
we first derive expressions of MSD for two types of Gaussian noise. 
First, we let $\xi(t)$  be the simple case of Gaussian white noise $\eta(t)$ 
with zero mean and covariance $C_\eta(t-s) = \left<\eta(t)\eta(s)\right> = \delta(t-s)$. 
Then 
(\ref{eq:DOfracLamgevin_00130}) 
becomes
\begin{align}
  \sigma^2(t) & = \int_0^t \left(G(u)\right)^2 du.
\label{eq:DOfracLamgevin_00140}
\end{align}
Evaluation of 
(\ref{eq:DOfracLamgevin_00140}) 
is in general complicated and it usually does not lead to a closed expression. 
One way to obtain a simpler expression for MSD is to impose the following condition on the Laplace transform of the covariance $\tilde{C}_\xi(s)$ of $\xi(t)$:
\begin{align}
  \tilde{G}(s)\tilde{C}(s) & = \frac{1}{s}.
\label{eq:DOfracLamgevin_00150}
\end{align}
The condition 
(\ref{eq:DOfracLamgevin_00150}) 
reduces 
(\ref{eq:DOfracLamgevin_00130}) 
to
\begin{align}
  \sigma^2(t) & = 2\int_0^t G(u)du
\label{eq:DOfracLamgevin_00160}
\end{align}
As we shall show in subsequent sections that Gaussian random noise with Laplace transform of its covariance satisfying condition 
(\ref{eq:DOfracLamgevin_00150}) 
not only facilitate the calculation of MSD, 
it also allows 
(\ref{eq:DOfracLamgevin_00060}) 
to model ultraslow diffusion which otherwise can not be done with the Gaussian white noise. 

Let us consider the properties of the Gaussian random noise $\xi(t)$
that satisfies condition
(\ref{eq:DOfracLamgevin_00150}).
From 
(\ref{eq:DOfracLamgevin_00150}) 
one gets
\begin{align}
  \tilde{C}(s) & = \frac{1}{s\tilde{G}(s)} = \frac{A(s)}{s} \nonumber \\
               &  = \int_0^1 \varphi(\alpha) s^{\alpha-1}d\alpha.
\label{eq:DOfracLamgevin_00170}
\end{align}
By noting that the inverse Laplace transform of $s^{\alpha-1}$ is $t^{-\alpha}/\Gamma(1-\alpha)$, 
one gets
\begin{align}
  C_\xi(t) & = \int_0^1 \frac{t^{-\alpha}}{\Gamma(1-\alpha)}\varphi(\alpha) d\alpha, \quad 0 \leq \alpha \leq 1.
\label{eq:DOfracLamgevin_00180}
\end{align}

Let us define the Gaussian random noise $\xi_\alpha(t)$ by
\begin{align}
  \left<\xi_\alpha(t)\right> & = 0
\label{eq:DOfracLamgevin_00190}
\end{align}
and
\begin{align}
  \left<\xi_\alpha(t)\xi_\alpha(s)\right> & = \frac{(t-s)^{-\alpha}}{\Gamma(1-\alpha)}
\label{eq:DOfracLamgevin_00200}
\end{align}
such that
\begin{align}
  C_\xi(t-s) & = \left<\xi(t)\xi(s)\right> 
           = \int_0^1 \int_0^1 \varphi(\alpha) \left<\xi_\alpha(t)\xi_\beta(s)\right> {d\alpha}{d\beta} \nonumber \\
         & = \int_0^1 \left<\xi_\alpha(t)\xi_\alpha(s)\right>\varphi(\alpha)  {d\alpha},
\label{eq:DOfracLamgevin_00210}
\end{align}
where we have used the orthogonality property of  
$\left<\xi_\alpha(t)\xi_\beta(s)\right>=\delta(\alpha-\beta) \left<\xi_\alpha(t)\xi_\alpha(s)\right>$.
Recall that the increment process of fractional Brownian motion or fractional Gaussian noise has the covariance  
$C_Ht^{2H-2}$, where $0 < H < 1$ is the Hurst index for the fractional Brownian motion, 
and $C_H=2H(2H-1)$.
By letting $\alpha=2-2H$, one can then identify $\xi_\alpha(t)$ with the fractional Gaussian noise 
(up to a multiplicative constant). 
Note that the fractional Gaussian noise is to be regarded as generalized derivatives of fractional Brownian motion or a generalized Gaussian random process 
\cite{Zinde_WalshPhillips03}.
We shall denote the distributed order fractional Gaussian noise with covariance 
(\ref{eq:DOfracLamgevin_00210})
by $\xi_\varphi(t)$, and it can be defined either as
\begin{subequations}
\label{eq:DOfracLamgevin_00220}
  \begin{align}
    \xi_\varphi(t) & = \int_0^1 \varphi(\alpha) \xi_\alpha(t) d\alpha,
    \label{eq:DOfracLamgevin_00220a}
  \end{align}
  or
  \begin{align}
    \xi_\varphi(t) & = \int_0^1 \sqrt{\varphi(\alpha)} \xi_\alpha(t) d\alpha,
    \label{eq:DOfracLamgevin_00220b}
  \end{align}
\end{subequations}
depending on the nature of the weight function $\varphi(\alpha)$ 
as we shall demonstrate in subsequent sections.

Before we end this section, it is necessary to point out some problems associated with 
the stochastic differential equation driven by fractional Gaussian noise. 
The question on whether a stochastic integral with respect to fractional Brownian motion leads to a well defined stochastic integral is a long standing problem which has attracted considerable 
attention (see 
\cite{Mushura08,BiaginiHuOksendal08}
and references therein).
Fractional Brownian motion is not a semimartingale if the Hurst index of the process $H \neq 1/2$, 
that is when the process is not a Brownian motion. 
As a result, the usual stochastic calculus of Ito can not be used to define the integrals with respect to fractional Brownian motion. 
Various methods such as Sokorohod-Stratonovich stochastic integrals, 
Malliavin calculus, and pathwise stochastic calculus have been suggested to overcome this problem 
(see references 
\cite{Mushura08,BiaginiHuOksendal08} 
for details). 
However, theory based on abstract integrals will encounter difficulty in physical interpretations in certain applications. 
Since our subsequent discussion 
deals with applications involving $1/2 < H < 1$, 
it is possible to consider the integrals with respect to fractional Brownian motion as the pathwise Riemann-Stieltjes integrals 
(see for example 
\cite{AzmoodehTikanmakiValkeila10} 
and references given there). 
In this way we can handle such integrals in a similar manner as ordinary integrals.

\section{Double Delta Function Distributed Order Fractional Langevin Equation}
\label{sec:DoubleOrder}

Fractional Langevin equation of double order results with the choice of the following weight function:
\begin{align}
  \varphi(\alpha) & = a_1\delta(\alpha - \alpha_1) + a_2\delta(\alpha - \alpha_2),                
\label{eq:DoubleOrder_00010}
\end{align}
where $1/2 \leq \alpha_1 < \alpha_2 \leq 1$.
The distributed order Langevin equation 
(\ref{eq:DOfracLamgevin_00060}) 
becomes
\begin{align}
  a_2D^{\alpha_2} x(t) + a_1D^{\alpha_1} x(t) & = \xi(t)
\label{eq:DoubleOrder_00020}
\end{align}
For both Riemann-Liouville and Caputo cases, one has
\begin{align}
  A(s) & = a_1 s^{\alpha_1} + a_2 s^{\alpha_2}
\label{eq:DoubleOrder_00030}
\end{align}
such that the Green function is given by the inverse Laplace transform of 
\begin{align}
  \frac{1}{A(s)} & = \frac{1}{a_1 s^{\alpha_1} + a_2 s^{\alpha_2}}
                  = \frac{1}{a_2}\frac{s^{-\alpha_1}}{s^{\alpha_2 - \alpha_1} + (a_1/a_2)}.
\label{eq:DoubleOrder_00040}
\end{align}
That is,
\begin{align}
  G(t) & = \frac{1}{a_2}t^{\alpha_2-1}E_{\alpha_2-\alpha_1,\alpha_2}\left(-\frac{a_1}{a_2}t^{\alpha_2-\alpha_1}\right),
\label{eq:DoubleOrder_00050}
\end{align}
where
\begin{align}
  E_{\mu,\nu}(z) & = \sum_{j=0}^\infty \frac{z^j}{\Gamma({\mu}j+\nu)}, \quad \mu > 0, \ \nu > 0,
\label{eq:DoubleOrder_00060}
\end{align}
is the Mittag-Leffler function 
\cite{Podlubny99}.

We consider first the case where the random noise is given by white noise $\eta(t)$, 
then the MSD of the process is given by
\begin{align}
  \sigma^2(t) & = \int_0^t\int_0^t G(u)\delta(u-v)G(v) dudv = \int_0^t \left[G(u)\right]^2 du \nonumber \\
              & = \frac{1}{a_2^2} \int_0^t u^{2(\alpha_2-1)}\left[
                E_{\alpha_2-\alpha_1,\alpha_2}\left(-\frac{a_1u^{\alpha_2-\alpha_1}}{a_2}\right)
              \right]^2 du
\label{eq:DoubleOrder_00070}
\end{align}
which can not be evaluated analytically. 
However, its asymptotic limits can be obtained and are given by
\begin{align}
  \sigma^2(t) & \sim \frac{t^{2\alpha_2-1}}{a_2^2(2\alpha_2-1)\left(\Gamma(\alpha_2)\right)^2},
                \quad \text{as} \ t \to 0  
\label{eq:DoubleOrder_00080}
\end{align}
and
\begin{align}
  \sigma^2(t) & \sim \frac{t^{2\alpha_1-1}}{a_1^2(2\alpha_1-1)\left(\Gamma(\alpha_1)\right)^2},
                \quad  \text{as} \ t \to \infty.
\label{eq:DoubleOrder_00090}
\end{align}

One can view the short time limit of MSD is obtained by ignoring the ${D^{\alpha_1}}x(t)$ in 
(\ref{eq:DoubleOrder_00020})
when $t \to 0$ and treated like $a_2{D^{\alpha_2}}x(t) = \xi(t)$.  
On the other hand, the long time limit of MSD is resulted when ${D^{\alpha_2}}x(t)$
   is neglected and 
(\ref{eq:DoubleOrder_00020})
reduces to $a_1{D^{\alpha_1}}x(t) = \xi(t)$ as $t \to \infty$. 
Therefore the process described by 
(\ref{eq:DoubleOrder_00020}) 
initially diffuses with scaling exponent $\alpha_2$, 
then it slows down to become a process with a smaller scaling exponent $\alpha_1$ as $t$ becomes very large. 
In other words, the process is asymptotically locally self-similar of order  
$\alpha_2 - 1/2$ with  $x(ct) = c^{\alpha_2 -1}x(t)$ for $c > 0$; 
and it is large-time asymptotically self-similar of order
$\alpha_1 - 1/2$ with  $x(ct) = c^{\alpha_1 -1}x(t)$.
Thus the resulting process describes retarding subdiffusion, 
which becomes more and more anomalous (or getting slower and slower) as time progresses. 
If we allow the limits of integration in 
(\ref{eq:introduction_00010})
to be $\beta_1 = 1$, $\beta_2 = 3/2$, and $1 < \alpha_1 < \alpha_2 <3/2$ in 
(\ref{eq:DoubleOrder_00010}), 
then the resulting process is retarding superdiffusion.
However, there is no way for one to obtain accelerating subdiffusion or superdiffusion based on 
(\ref{eq:DoubleOrder_00020}).

Next we consider the case where the random noise $\xi(t)$ in 
(\ref{eq:DOfracLamgevin_00060}) 
is given by the distributed order fractional Gaussian noise 
\begin{align}
  \xi_\varphi(t) & = \int_0^1 \xi_\alpha(t)\varphi(\alpha) d\alpha
                 =  \sqrt{a_1}\xi_{\alpha_1}(t) + \sqrt{a_2}\xi_{\alpha_2}(t)
\label{eq:DoubleOrder_00100}
\end{align} 
with covariance given by
\begin{align}
  C_\xi(t) & = a_1\frac{t^{-\alpha_1}}{\Gamma(1-\alpha_1)} + a_2\frac{t^{-\alpha_2}}{\Gamma(1-\alpha_2)}
\label{eq:DoubleOrder_00110}
\end{align}
The random noise $\xi(t)$ can be regarded as the sum of two fractional Gaussian noises correspond to Hurst indices
$H_i=1-\frac{\alpha_i}{2}$, $i=1,2$

Stochastic process associated with the distributed order fractional Langevin equation 
(\ref{eq:DoubleOrder_00020}) 
with the above Gaussian random noise $\xi_\varphi(t)$ has the following MSD:
\begin{align}
  \sigma^2(t) & = \frac{2}{a_2} \int_0^t u^{\alpha_2-1} 
                  E_{\alpha_2-\alpha_1,\alpha_2}\left(-\frac{a_1}{a_2}u^{\alpha_2-\alpha_1}\right)du \nonumber \\
              & =  \frac{2}{a_2}t^{\alpha_2}E_{\alpha_2-\alpha_1,\alpha_2}\left(-\frac{a_1}{a_2}t^{\alpha_2-\alpha_1}\right)
\label{eq:DoubleOrder_00120}
\end{align}
The long and short time limits are then given by
\begin{align}
  \sigma^2(t) \sim \frac{2}{a_2\Gamma(\alpha_2+1)}t^{\alpha_2} \quad \text{as} \ t \to 0
\label{eq:DoubleOrder_00130}
\end{align}
and
\begin{align}
  \sigma^2(t) \sim \frac{2}{a_1\Gamma(\alpha_1+1)}t^{\alpha_1} \quad \text{as} \ t \to \infty
\label{eq:DoubleOrder_00140}
\end{align}
Thus we see that the solution of 
(\ref{eq:DoubleOrder_00020}) 
with Gaussian white noise and Gaussian distributed fractional noise 
both lead to power law type of MSD with different exponents for the short and long time limits of MSD.  

\begin{figure}[ht]
  \centering
  \subfigure[\label{fig:DoubleOrder_01a}]{
    \includegraphics[width=6.5cm]{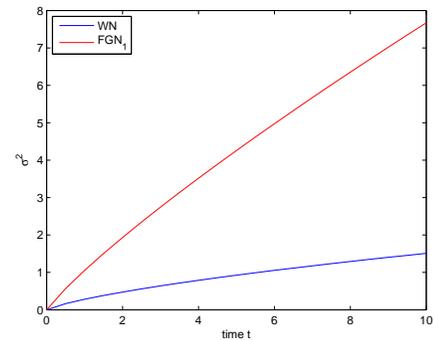}} \\
  \subfigure[\label{fig:DoubleOrder_01b}]{
    \includegraphics[width=6.5cm]{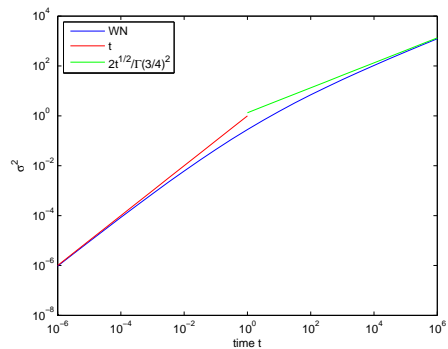}} 
  \subfigure[\label{fig:DoubleOrder_01c}]{
    \includegraphics[width=6.5cm]{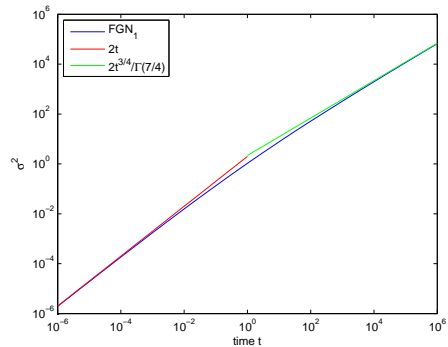}}
  \caption{\label{fig:DoubleOrder_01}
    MSD of double delta-function distributed order process with $a_1=a_2=1$, $\alpha_2=1$ and $\alpha_1 =3/4$. 
    (a) MSD associated with distributed order equation with white noise (WN) and 
         double delta function distributed order fractional Gaussian noise ($\rm FGN_1$); 
    (b) log-log plot of MSD associated with double-order distributed order fractional Langevin equation with WN; 
    (c) log-log plot of MSD associated with double-order distributed order fractional Langevin equation with $\rm FGN_1$.
  }
\end{figure}

Now we want to see whether the processes in the above examples can be used to describe certain transport phenomena in physical systems.
Clearly, such a process must not have unique characteristic scaling exponent, 
instead it has piecewise scaling exponents. 
One such physical process which has different scaling for short and long time behavior of MSD is single-file diffusion. 
Recall that in single-file diffusion, 
the particles are geometrically constraint to move in a line and unable to overtake 
\cite{KalgesRadonsSokolov08,KargerRuthven92,LimTeo09}.
For very short times the MSD of the diffusing particles varies with time, 
just like in normal diffusion.
The Brownian particles in confined geometry such as nanopores or nanochannels can not alter their relative ordering, 
therefore the subsequent motion of each particle is always constrained by the same two neighboring particles. 
Thus, in the long time limit this effect of caging slows down the diffusion, 
and changes the MSD from linear growth to one that varies with $\sqrt{t}$. 
If white noise is used in 
(\ref{eq:DoubleOrder_00020}), 
when $\alpha_2 = 1$ and $2\alpha_1-1 = 1/2$, or $\alpha_1 = 3/4$, 
the limits 
(\ref{eq:DoubleOrder_00080}) 
and 
(\ref{eq:DoubleOrder_00090}) 
give the correct asymptotic behavior of the MSD for single-file diffusion 
\cite{KalgesRadonsSokolov08,KargerRuthven92,LimTeo09}.
On the other hand, if we use distributed order Gaussian fractional noise in 
(\ref{eq:DoubleOrder_00020}),
then for $\alpha_2=1$ and $\alpha_1 = 1/2$, 
one obtains the correct short and long time limits for the MSD of single-file diffusion. 
Figure \ref{fig:DoubleOrder_01a} shows the comparison of MSD obtained 
using Gaussian white noise and distributed order fractional Gaussian noise, 
and Figures \ref{fig:DoubleOrder_01b} and \ref{fig:DoubleOrder_01c} show the short and long time limit of MSD for these two cases.

Recall that Brownian motion is related to white noise 
(in the sense of generalized function) 
by the following free Langevin equation $Dx(t) = \eta(t)$. 
If we regard the process as Brownian motion experiences a retardation due to the confined geometry,
then with $\alpha_2=1$ and $\alpha_1 = 3/4$ such that the second term with the fractional derivative in 
(\ref{eq:DoubleOrder_00020}) 
can be regarded as a damping or retarding term that slows down the Brownian motion 
so that the motion which begins as normal diffusion becomes single-file subdiffusion after a long time. 
Since we only use 
(\ref{eq:DoubleOrder_00020}) 
to describe the asymptotic properties of single-file diffusion, 
it does not give a unique description of single-file diffusion.  
It is necessary to consider the behavior of single-file diffusion at intermediate times to see whether it also agrees with the corresponding description given by the distributed order fractional Langevin equation 
(\ref{eq:DoubleOrder_00020}).

Here we would like to remark that a similar asymptotic behavior for the MSD can also be obtained by using fractional time diffusion equation of distributed order 
\cite{UmarovSteinberg09,LimTeo09,Caputo03}
\begin{align}
  \int_0^1 \varphi(\alpha) D_t^\alpha W(x,t) d\alpha & = \frac{\partial^2 W(x,t)}{\partial{x}^2}, 
\label{eq:DoubleOrder_00150}
\end{align}
where $x \in \mathbb{R}, t \geq 0$ and $W(x,t)$  is the probability distribution function. 
Using the same weighing function $\varphi(\alpha)$ as in 
(\ref{eq:DoubleOrder_00020}) 
and $D^\alpha$ as Caputo fractional derivative one gets
  \begin{align}
    a_1\frac{\partial^{\alpha_1} W(x,t)}{\partial{t}^{\alpha_1}} + a_2\frac{\partial^{\alpha_2} W(x,t)}{\partial{t}^{\alpha_2}} 
 & = \frac{\partial^2W(x,t)}{\partial{x}^2}. 
    \label{eq:DoubleOrder_00160}
  \end{align}
Using the initial condition $W(x,0^{+}) = \delta(x)$,
the solution of 
(\ref{eq:DoubleOrder_00160}) 
is a diffusion process with variance having asymptotic behavior  
$\sigma^2 \sim t^{\alpha_1}$ 
and  
$\sigma^2 \sim t^{\alpha_2}$
as large time and short time limit respectively.
However, if the fractional time derivative of Riemann-Liouville 
type is used in (\ref{eq:DoubleOrder_00150}), 
then the asymptotic behavior is opposite to that for the Caputo derivative with the smaller exponent $\alpha_1$ 
dominates at small time and larger exponent $\alpha_2$ dominates at large time. 
Thus, in contrast to 
(\ref{eq:DoubleOrder_00020}), 
it is possible for to obtain accelerating subdiffusion based Riemann-Liouville version of 
(\ref{eq:DoubleOrder_00150}).
Another major difference between these two approaches is that the description based on the distributed order time fractional diffusion equation 
(\ref{eq:DoubleOrder_00150})
is a non-Gaussian model, whereas the distributed order fractional Langevin equation 
(\ref{eq:DoubleOrder_00020}) 
is a Gaussian one.
We would like to remark that there also exists an effective Fokker-Planck equation which leads to a similar results, and it provides a Gaussian model for single-file diffusion 
\cite{EabLim10a}.

\section{Uniformly Distributed Order Fractional Langevin Equation}
\label{sec:uniform}

There exists a class of strongly anomalous diffusion with the long time limit of its MSD decays logarithmically as
$\left(\log{t}\right)^\kappa$, $\kappa > 0$. 
Such ultraslow diffusion occurs in Sinai diffusion of a particle in a one-dimensional quenched random energy landscape
\cite{Sinai82,ChaveGuitter99}, 
in charged polymers 
\cite{SchiesselSokolovBlumen97}, 
motion in aperiodic environments 
\cite{Igloi99}, 
in a class of iterated maps 
\cite{DragerKlafter00}, 
in area preserving parabolic map 
\cite{ProsenZnidaric21} 
and charged tracer particle on a two-dimension lattice 
\cite{BenichouOshanin02}, 
etc. 
It has been shown that uniformly distributed order time fractional diffusion equation can be used to model the ultraslow diffusion 
\cite{%
ChechkinGorenfloSokolov02,%
ChechkinGSG02,%
ChechkinKlafterSokolov03,%
Naber04,%
GorenfloMainardi05,%
MeerschaertScheffler06,%
Hanyga07,%
Kochubei08,%
MainardiMPG08%
}.

In this section we want to consider uniformly distributed order fractional Langevin equation to see whether it can describe ultraslow diffusion. 
For uniform distributed order the weight function is $\varphi(\alpha)=1$, $0 \leq \alpha \leq 1$. 
Now 
(\ref{eq:DOfracLamgevin_00060}) 
becomes
\begin{align}
  \int_0^1 {D^\alpha}x(t)d\alpha & = \xi(t)
\label{eq:uniform_00010}
\end{align}
which gives the Laplace transform of its Green function as
\begin{align}
  \tilde{G}(s) & = \frac{1}{A(s)} = \left[\int_0^1{s^\alpha}d\alpha\right]^{-1} = \frac{\log{s}}{s-1}
\label{eq:uniform_00020}
\end{align}
By taking the inverse Laplace transform one gets
\begin{align}
  G(t) & = e^t\mathbf{E}_1(t),
\label{eq:uniform_00030}
\end{align}
with
\begin{align}
  \mathbf{E}_1(t) & = - \gamma - \log{t} + \mathbf{Ein}(t),
\label{eq:uniform_00040}
\end{align}
$\mathbf{E}_1(t)$  is the exponential integral function given by 
\cite{AbramowitzStegun71}
\begin{align}
   \mathbf{E}_1(z) & = \int_{z}^\infty \frac{e^y}{y}dy
\label{eq:uniform_00050}
\end{align}
and
\begin{align}
  \mathbf{Ein}(t) & = \int_0^t \frac{1-e^{-u}}{u} = \sum_{k=1}^\infty \frac{(-1)^{k+1}t^k}{kk!}.
\label{eq:uniform_00060}
\end{align}

\begin{figure}[ht]
  \centering
  \subfigure[\label{fig:uniform_01a}]{
    \includegraphics[width=6.5cm]{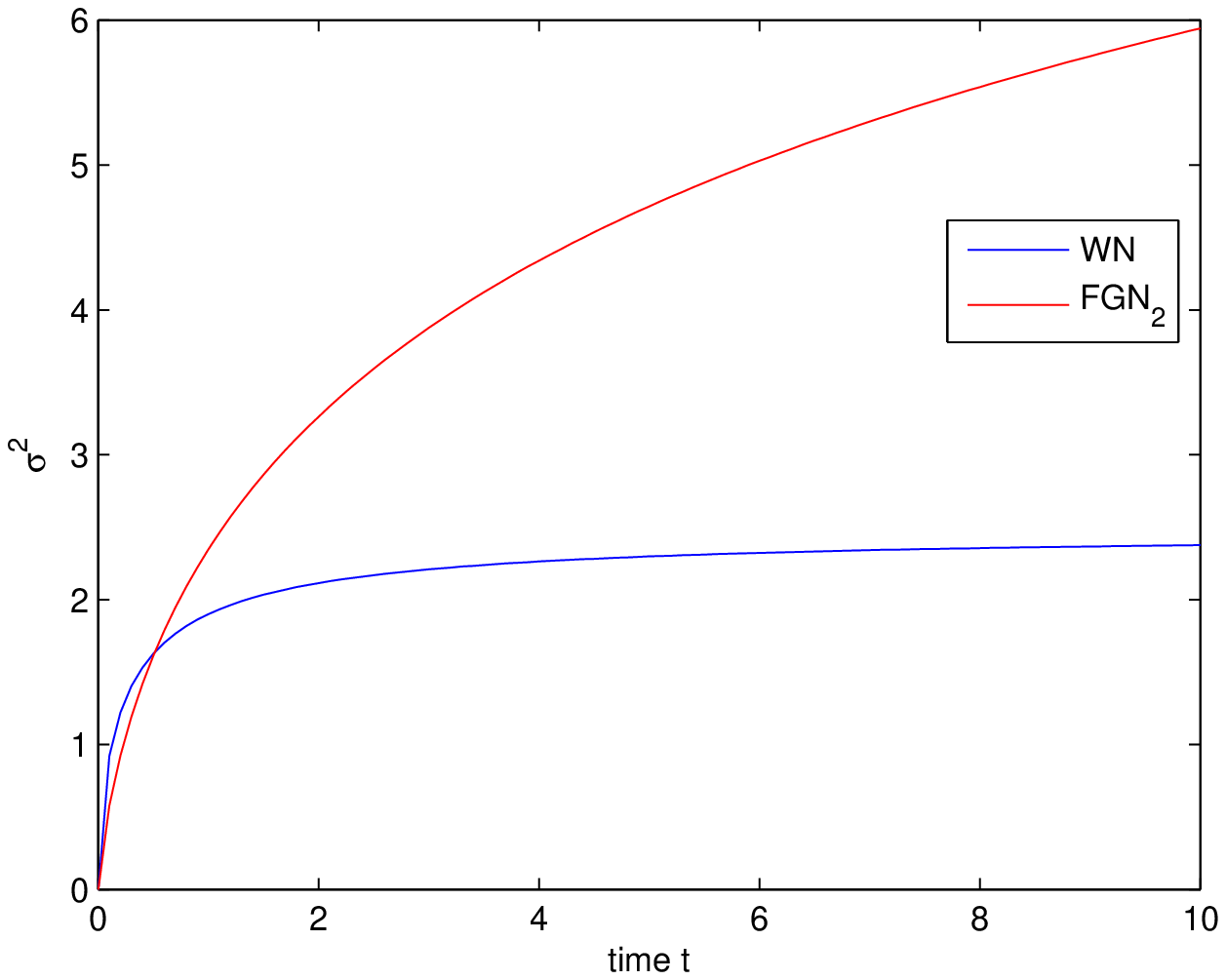}} \\
  \subfigure[\label{fig:uniform_01b}]{
    \includegraphics[width=6.5cm]{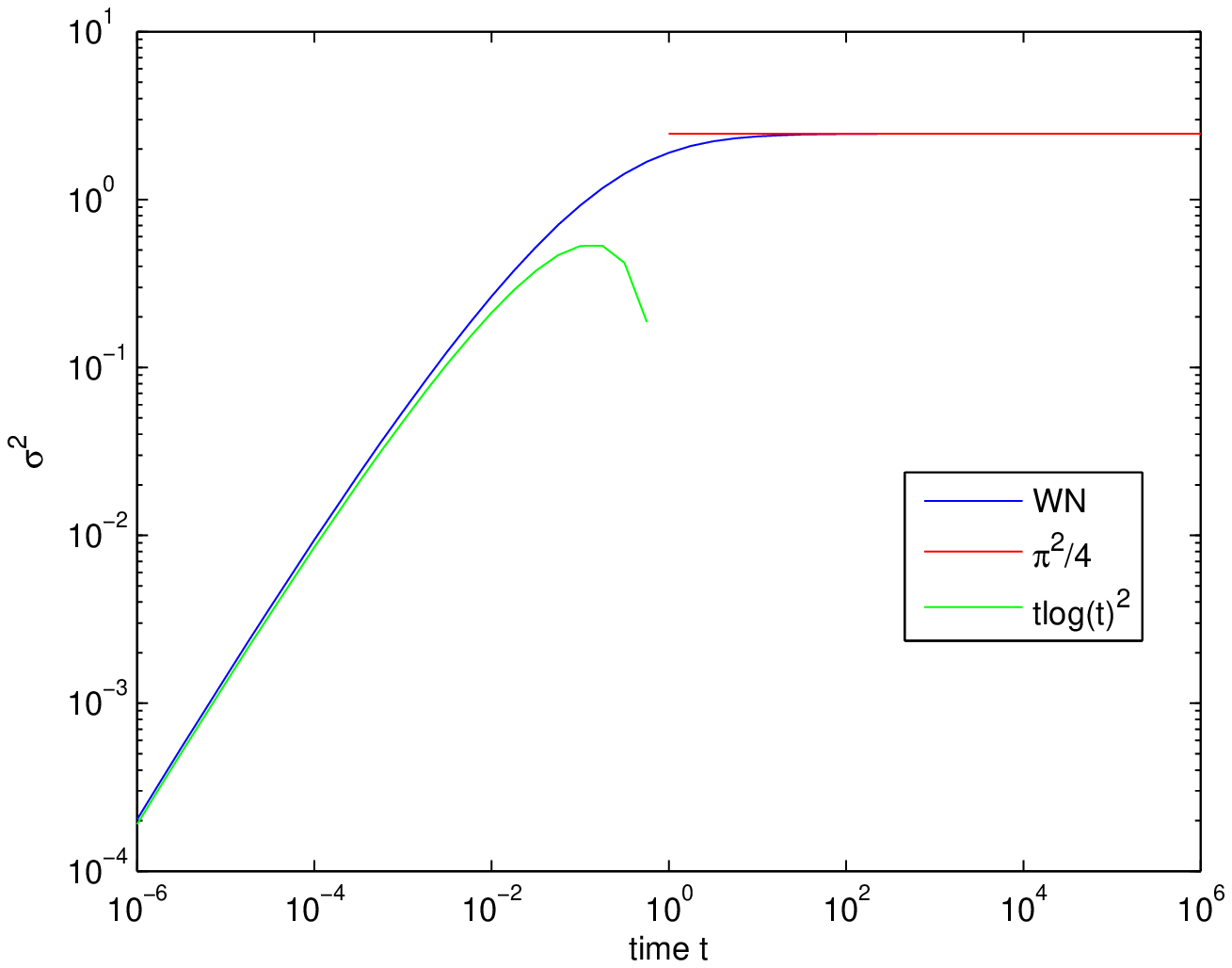}}
  \subfigure[\label{fig:uniform_01c}]{
    \includegraphics[width=6.5cm]{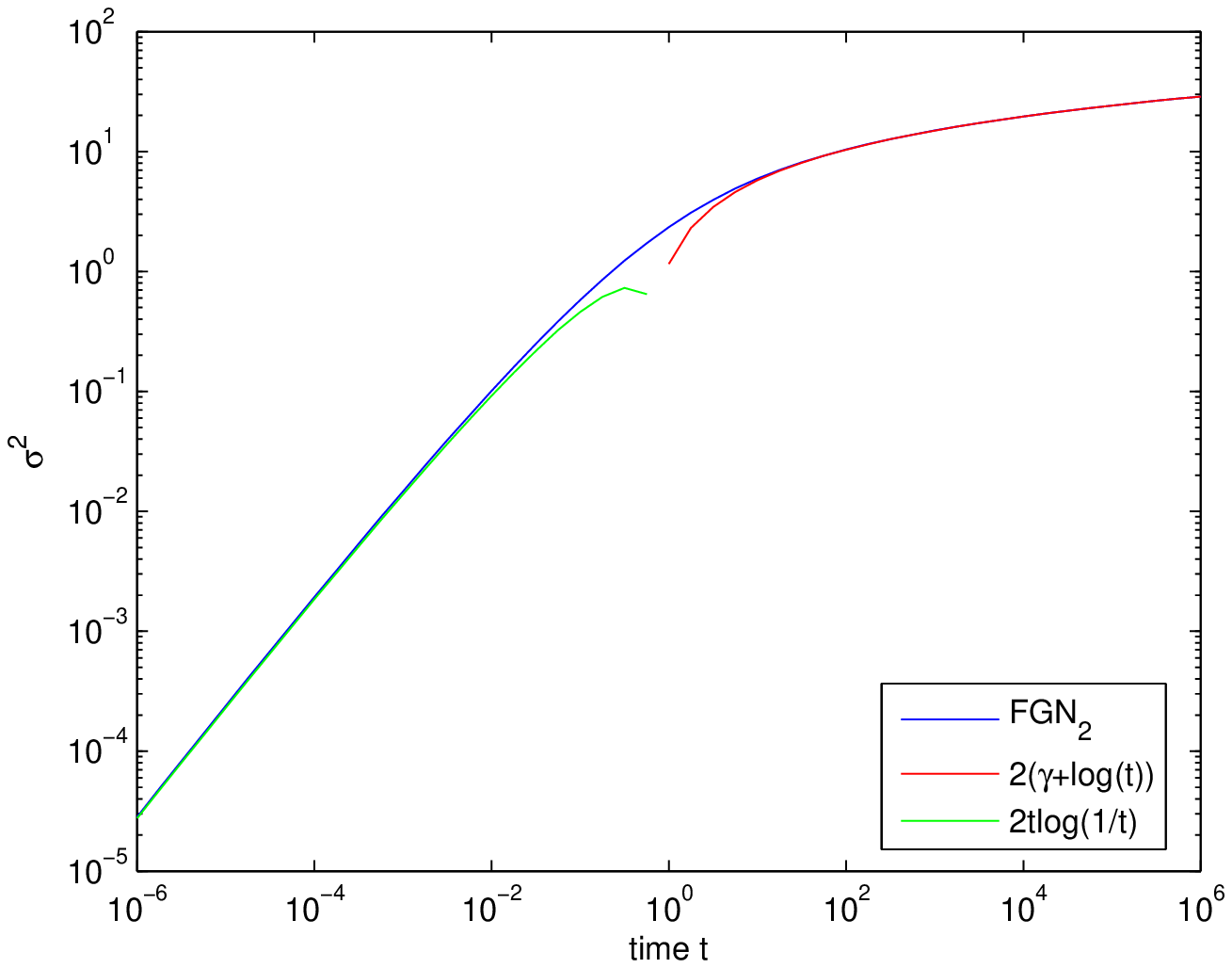}}
  \caption{\label{fig:uniform_01}
    MSD of the uniformly distributed process 
corresponds to 
the uniformly distributed order in the range $\alpha \in [0,1]$:
    (a) MSD of the process associated with distributed uniformly order fractional Langevin equation with white noise (WN) and uniformly distributed fractional Gaussian noise ($\rm FGN_2$); 
    (b) log-log plot of MSD associated with uniformly distributed order fractional Langevin equation with WN; 
    (c) log-log plot of MSD associated with the distributed order fractional Langevin equation with $\rm FGN_2$.
  }
\end{figure}

In the case where the random noise $\xi(t)$ is white noise, the MSD is given by
\begin{align}
    \sigma^2(t) & = \int_0^t \left[e^u\mathbf{E}_1(u)\right]^2 du,
\label{eq:uniform_00070}
\end{align}
which can not be evaluated analytically. 
In order to study the asymptotic behavior of the MSD, 
we consider the upper and lower bound of the exponential integral function 
(see 
\cite{AbramowitzStegun71},
\#5.1.20):
\begin{gather}
  \frac{1}{2}\log\left(1+\frac{2}{t}\right) < e^t\mathbf{E}_1(t) < \log\left(1+\frac{1}{t}\right).
\label{eq:uniform_00080}
\end{gather}
From 
(\ref{eq:uniform_00070})
the upper bound of the variance is 
\begin{align}
  U(t) & = \int_0^t\left[\log\left(1+\frac{1}{u}\right)\right]^2 du  \nonumber \\
       &  = \frac{\pi^2}{3} + t\left[\log\left(\frac{t}{1+t}\right)\right]^2 
          - 2\sum_{n=1}^\infty \frac{1}{n^2(1+t)}.
  \label{eq:uniform_00090}
\end{align}
The lower bound is
\begin{align}
  L(t) & = \frac{1}{4}\int_0^t\left[\log\left(1+\frac{2}{u}\right)\right]^2 du \nonumber \\
       &  = \frac{1}{2}\int_0^{t/2}\left[\log\left(1+\frac{1}{u}\right)\right]^2 du 
         = \frac{1}{2}U\left(\frac{t}{2}\right)
\label{eq:uniform_00100}
\end{align}
When $t \to 0$, the summation term in 
(\ref{eq:uniform_00090}) 
tends to zeta function $2\zeta(2) = \pi^2/3$,  
which cancels with the first term in the equation. 
Therefore one gets 
\begin{align}
    U(t) \underset{t \to 0}{\sim} t\left[\log(t)\right]^2
\label{eq:uniform_00110}
\end{align}
and
\begin{align}
    L(t) \underset{t \to 0}{\sim} \frac{t}{4}\left[\log(t)\right]^2.
\label{eq:uniform_00120}
\end{align}
Thus, the short-time limit of the MSD is given by
\begin{align}
  \sigma^2(t) \underset{t \to 0}{\sim} c_1t\left[\log(t)\right]^2 = c_1t\left[\log(1/t)\right]^2
\label{eq:uniform_00130}
\end{align}
where $1/4 \leq c_1 \leq 1$. 
From numerical simulations, 
we get $c_1 = 1$ as shown in 
Figure \ref{fig:uniform_01b}. 
For the large-time limit of the MSD, 
one notes that the summation term in 
(\ref{eq:uniform_00090}) 
tends to zero as $t \to \infty$, 
and the second term of 
(\ref{eq:uniform_00090}) 
becomes
\begin{align}
  t\left[\log\left(\frac{t}{1+t}\right)\right]^2  & = t\left[-\frac{1}{t}+\frac{1}{2t^2} - \cdots\right]^2 \nonumber \\
                  & \underset{t\to\infty}{\sim} \frac{1}{t} + \mathcal{O}\left(\frac{1}{t^2}\right).
\label{eq:uniform_00140}
\end{align}
Thus,
\begin{align}
  U(t) \underset{t \to \infty}{\sim} \frac{\pi^2}{3} 
\label{eq:uniform_00160}
\end{align}
and 
\begin{align}
  L(t) \underset{t \to \infty}{\sim} \frac{\pi^2}{6}.
\label{eq:uniform_00170}
\end{align}
Therefore the MSD approaches a constant for sufficient large time,
\begin{align}
  \sigma^2(t) \underset{t \to \infty}{\sim} c_2\frac{\pi^2}{6},
\label{eq:uniform_00180}
\end{align}
where $c_2 = 1.5$ can be obtained graphically. 
One may retain the time-dependent term in the MSD for large (but not too large) time:
\begin{align}
  \sigma^2(t) \underset{t \to \infty}{\sim} \frac{\pi^2}{4} + \frac{1}{t}
\label{eq:uniform_00190}
\end{align}

Here we have a motion that begins as a non-stationary process and becomes a stationary one after sufficiently long time. 
In fact, at small times the diffusion is anomalous diffusion of slightly superdiffusive type. 
In other words, the process goes to zero slower than the normal diffusion due to the $\log(1/t)$ term as $t \to 0$. 
However, at large times it tends to a stationary process with a constant variance. 
One can interprete the long time behavior in the following way. 
The uniformly distributed derivative is the derivative
$D^\alpha$ integrated over the range $\alpha=0$ to $\alpha=1$.
As $t \to \infty$, one would expect the dominant term will be from $\alpha=0$, which will result
$x(t) = \eta(t)$, a white noise process with constant variance.
In comparison, 
for distributed order time fractional diffusion equation  
(\ref{eq:DoubleOrder_00150}) 
with uniformly distributed order, the associated process at short time behaves somewhat superdiffusive, 
with variance $\sim t\log(1/t)$; 
and for long time limit the diffusion process becomes ultraslow with variance $\sim \log(t)$. 

Now instead of using white noise in
(\ref{eq:uniform_00010}),
we let the random noise $\xi(t)$  
be the uniformly distributed fractional Gaussian noise, 
\begin{align} 
  \xi_\varphi(t) & =  \int_0^1 \xi_\alpha(t)d\alpha,
\label{eq:uniform_00200}
\end{align}
where the weight function is given by $\varphi(\alpha) = 1$. 
From 
(\ref{eq:uniform_00200})
one gets the covariance of $\xi_\varphi(t)$ as
\begin{align}
  C_\varphi(t -s) & = \left<\xi_\varphi(t)\xi_\varphi(s)\right> = \int_0^1 \frac{(t-s)^{-\alpha}}{\Gamma(1-\alpha)}d\alpha.
\label{eq:uniform_00210}
\end{align}
The MSD is then given by
\begin{align}
  \sigma^2(t) & = 2\int_0^t G(u) du = 2 \int_0^t e^{u}\mathbf{E}_1(u)du \nonumber \\
              & = 2\left(e^t\mathbf{E}_1(t)+ \gamma + \log(t)\right)
\label{eq:uniform_00220}
\end{align}
where $\gamma$ is the Euler's constant.
Substituting (\ref{eq:uniform_00040}) into (\ref{eq:uniform_00220}) gives
\begin{align}
  \sigma^2(t) & = 2e^t\mathbf{Ein}(t) + 2\left(1-e^t\right)\left(\gamma + \log(t)\right) 
\label{eq:uniform_00230}
\end{align}
Hence, the short time limit of the MSD is given by
\begin{align}
    \sigma^2(t)  & \underset{t \to 0}{\sim} 2t\log\left(\frac{1}{t}\right).
\label{eq:uniform_00250}
\end{align}

Using (\ref{eq:uniform_00080}) in (\ref{eq:uniform_00220}), one gets the upper bound and lower bounds of MSD as
  \begin{subequations}
    \label{eq:uniform_00260}
    \begin{align}
      H(t) & = 2\left[\log\left(\frac{1+t}{t}\right)+\gamma +\log(t)\right] \nonumber \\
           & = 2\left[\log(1+t) +\gamma\right]
      \label{eq:uniform_00260a}\\
      L(t) & = 2\left[\frac{1}{2}\log\left(\frac{2+t}{t}\right)+\gamma +\log(t)\right] \nonumber \\
           & = 2\left[\frac{1}{2}\log\left[t(2+t)\right] + \gamma\right]
      \label{eq:uniform_00260b}
    \end{align}
  \end{subequations}
Both upper and lower bound are approach to the same assymptotic function, thus we have
\begin{align}
  \sigma^2(t) \underset{t \to \infty}{\sim} 2\log(t) +2\gamma
\label{eq:uniform_00261}
\end{align}
Therefore, the large time limit of the MSD is
\begin{align}
  \sigma^2(t) \underset{t \to \infty}{\sim} 2\log(t)
\label{eq:uniform_00270}
\end{align}
Note the constant term $\gamma \approx 0.57722$ can not be neglected in practice since even for
$t=e^{10}$, $\sigma^2(t) \approx 2(10+0.57722)$ which shows a 6 \% contribution from the Euler's constant.
The above asymptotic behavior of the MSD shows that the diffusion is ultraslow at very large times, 
and it becomes slightly superdiffusive at small times. 

Figure \ref{fig:uniform_01a} shows comparison of the MSD of the stochastic process associated with uniformly distributed order Langevin equations with Gaussian white noise and uniformly distributed fractional Gaussian noise. 
In Figures \ref{fig:uniform_01b} and \ref{fig:uniform_01c}, 
the short and long time limit of MSD for these two cases is demonstrated.

\section{Power Law Distributed Order Fractional Langevin Equation}
\label{sec:powerlaw}

In order to describe ultraslow diffusion processes with large time limit of MSD varies as
$(\log{t})^\nu$, $\nu > 0$,
it is necessary to consider power-law distributed order fractional Langevin equation with weighing function
$\nu\alpha^{\nu-1}$, $\nu > 0$. 
(\ref{eq:DOfracLamgevin_00060}) 
now becomes
\begin{align}
  \nu\int_0^1 \alpha^{\nu-1}{D^\alpha}x(t)d\alpha & = \xi(t).
\label{eq:powerlaw_00010}
\end{align}
Laplace transform of the Green function $\tilde{G}(s)$ of 
(\ref{eq:powerlaw_00010}) 
is the inverse of
\begin{align}
  A(s) & =\nu\int_0^1 \alpha^{\nu-1}{s^\alpha} d\alpha.
\label{eq:powerlaw_00020}
\end{align}
For $0 < s < 1$ or $ -\infty < \log{s} < 0$, one obtains
\begin{align}
  \tilde{G}(s) & = \frac{(-\log{s})^\nu}{\nu\gamma(\nu,-\log{s})}.
\label{eq:powerlaw_00030}
\end{align}
Since $\gamma(\nu,z) \sim \Gamma(\nu)$ as $z \to \infty$, thus for small $s$ or large $-\log{s}$,
\begin{align}
  \tilde{G}(s) \sim \frac{\left(\log(1/s)\right)^\nu}{\nu\Gamma(\nu)} \quad \text{as} \ s \to 0.
\label{eq:powerlaw_00040}
\end{align}
If we assume the Gaussian random noise in 
(\ref{eq:powerlaw_00010}) 
is power-law distributed order Gaussian fractional noise, 
\begin{align}
    \xi_\varphi(t) & = \nu \int_0^1 \alpha^{\nu-1} \xi_\alpha(t)d\alpha
\label{eq:powerlaw_00050}
\end{align}
then the Laplace transform of the MSD is given by $\widetilde{\sigma}^2(s) = 2\tilde{G}(s)/s$, 
which can be verified as a slowly varying function 
\cite{Kochubei08,BinghamGoldieTeugels87}.
Now applying Tauberian theorem 
\cite{Feller71,Korevaar04},
which allows the long and short time asymptotic limits of a function $f(t)$ to be obtained from the Laplace transform 
$\tilde{f}(s)$ for $s$ near origin and infinity respectively 
[see for example, page 445, reference 
\cite{Feller71}
). 
Thus from 
(\ref{eq:powerlaw_00040}),
one gets
\begin{align}
  \sigma^2(t) & \sim 2 \frac{(\log{t})^\nu}{\Gamma(\nu+1)} \quad \text{as} \ t \to \infty.
\label{eq:powerlaw_00060}
\end{align}
Similarly, the short time limit for the MSD can be obtained from the large $s$ limit of $\widetilde{\sigma}^2(s)$ given by
\begin{align}
  \widetilde{\sigma}^2(s) & \sim  \frac{\log{s}}{{\nu}s} \quad \text{as} \ s \to \infty.
\label{eq:powerlaw_00070}
\end{align}
From this we get the short time limit for the MSD as
\begin{align}
    \sigma^2(t) & \sim 2 \frac{t\log(1/t)}{\nu} \quad \text{as} \ t \to 0.
\label{eq:powerlaw_00080}
\end{align}
Thus, the distributed order fractional Langevin equation with power law weight function provides a way to describe the kinetics of the ultraslow diffusion such as Sinai diffusion with $\nu =4$ for particles moves in a quenched random field, 
and transport of hooked polyampholytes (heteropolymers which carry both positive and negative charges) with $\nu = 4/3$. 

\section{Concluding Remarks}
\label{sec:conclude}

We have shown that distributed order fractional Langevin equation provides a mathematical model for anomalous diffusion which does not have a unique scaling exponent. 
It is interesting to note that the expression for MSD acquires  a more simple form
 if white noise in the distributed order fractional Langevin equation is replaced by distributed order fractional Gaussian noise.
The solutions of distributed order fractional Langevin equations have MSD which describe retarding subdiffusion such as in single-file diffusion, 
and ultraslow diffusion with logarithmic growth. 
To a large extent, 
the results obtained are similar to that from time-fractional diffusion equation of distributed order, 
except for one main difference that the MSD for the Langevin case has same properties for both Riemann-Liouville and Caputo distributed derivatives, 
whereas in the fractional diffusion equation Riemann-Liouville and Caputo distributed derivatives lead to MSD with different behavior.  
In addition, distributed order time fractional diffusion equation result in a non-Gaussian process, 
whereas in the process obtained from the corresponding Langevin equation is Gaussian.

Possible direct generalizations of our study are extensions of free Langevin equation to fractional Langevin equation and fractional generalized Langevin equation of distributed order. 
However, one notes that "frictional terms appear in the free Langevin equation once the usual fractional derivative is replaced by the distributed order fractional derivative.
For example, free fractional Langevin equation of distributed order with the weight function
$\varphi(\alpha) = a_1\delta(\alpha-\alpha_1) + a_2\delta(\alpha-\alpha_2) + \lambda\delta(\alpha)$,
$\lambda > 0$
will become  
$a_2D^{\alpha_2}x(t) + a_1D^{\alpha_1}x(t) + {\lambda}x(t) = \xi(t)$. 
Hence frictional terms can appear in free Langevin equation 
when the time derivative is replaced by distributed order time derivative with certain weight function.
The solution is complex in such a case as the Green function involves sum of Wright functions 
\cite{KilbasSrivastavaTrujillo06},
and the associated MSD is even be more complicated to compute. 
In the case of generalized fractional Langevin equation of distributed order 
${D_{(\varphi)}}x(t) + \int_0^t\gamma(t-u)x(u)du = \xi(t)$,
where $\gamma(t)$ is the frictional kernel, 
one would expect it to be mathematically more involved, though the assumption of fluctuation-dissipation theorem may help to simplify the situation somewhat.  
The main difficulty is in the evaluation of the inverse Laplace transform for the Green function 
$\tilde{G}(s) = \tilde{G}_{\varphi}(s)\left[1+\tilde{G}_{\varphi}(s)\tilde{\gamma}(s)\right]^{-1}$,
where $\tilde{G}_{\varphi}(s)$ 
denotes Laplace transform of Green function for ${D_{(\varphi)}}x(t)=0$.  
All these generalizations are not only computationally complex and mathematically intractable, 
they may not lead to very interesting results. 
Our study shows that the simpler fractional Langevin equation of distributed order seem to be adequate 
for describing the kinetics of the types of diffusion under consideration, 
hence it provides a viable alternative to the time fractional diffusion equation of distributed order.


\providecommand{\noopsort}[1]{}\providecommand{\singleletter}[1]{#1}%

\end{document}